\documentclass[]{raa}  

\usepackage{graphicx,times}
\usepackage{natbib}
\usepackage{amssymb,amsmath}
\bibpunct{(}{)}{;}{a}{}{,}

\newcommand\msun{\,\mathrm{M}_\odot}

\newcommand\too{\,\mathrm{to}\,}

\usepackage[pagebackref=true]{hyperref}

\begin{document}

   \title{A new X-ray tidal disruption event candidate with fast variability}

 \volnopage{ {\bf 20XX} Vol.\ {\bf X} No. {\bf XX}, 000--000}
   \setcounter{page}{1}

   \author{J. Hampel
   \inst{1}\thanks{Deceased September 30, 2017.}, S. Komossa\inst{2,3}, J. Greiner\inst{4}, T.H. Reiprich\inst{1}, M. Freyberg\inst{4}, T. Erben\inst{1}
   }
   
   \institute{Argelander-Institut f\"ur Astronomie, Auf dem H\"ugel 71, 53121 Bonn, Germany \\  
 \and
Max-Planck Institut f\"ur Radioastronomie, Auf dem H\"ugel 69, 53121 Bonn, Germany; {\it skomossa@mpifr.de} \\
\and
National Astronomical Observatories, Chinese Academy of Science, 20A Datun Road, Chaoyang District, Beijing 100012, China \\ 	
\and 
Max-Planck Institute f\"ur Extraterrestrische Physik, Gießenbachstrasse 1, 85748 Garching, Germany \\
\vs \no
   {\small Received 2021 December 4; accepted 2022 Month Day}
}

\abstract{
During a close encounter between a star and a supermassive black hole, the star can get disrupted by the black hole's tidal forces, resulting in a tidal disruption event (TDE). The accretion of the star's material onto the black hole produces strong emission in different wavelength regimes. Here we report the discovery with ROSAT of an X-ray-selected transient source in an optically non-active galaxy. At the location RA: $13^\mathrm{h}31^\mathrm{m}57.66^\mathrm{s}$ and Dec: $-32^\circ43'19.7''$ a sudden rise in X-ray luminosity by a factor of $8$ within 8 days
has been observed. Additionally, a very soft X-ray spectrum with 
a black-body temperature $kT$=0.1 keV and 
a peak luminosity of at least $1 \times 10^{43}\,\mathrm{erg/s}$ suggest a TDE interpretation,
and the observed properties are very similar to previously identified soft X-ray (ROSAT) TDEs. 
An optical spectrum taken of the galaxy at the position of RXJ133157.6–324319.7 
six years after the X-ray outburst does not show any emission lines as would be expected from a persistent active galactic nucleus (AGN). The redshift of the galaxy is determined to be 0.051 based on absorption lines. It is therefore likely a member of the galaxy cluster Abell 3560.
The rise in X-ray luminosity happens within 8 days and thus appears to be fast for such an event. No X-ray emission was detected 170 days before and 165 days after the event, and none was detected 25 years later with  
the
Neil Gehrels Swift Observatory. The change in X-ray luminosity is at least a factor of 40.
}

   \authorrunning{J. Hampel et al. }            %author_head in even pages
   \titlerunning{A new X-ray TDE}  % title_head in odd pages
   \maketitle

%________________________________________________ sections below
% 
\section{Introduction}           %% first-level sections will be auto-capitalized
\label{sect:intro}

   Tidal disruption events (TDEs) represent the disruption of a star due to a close encounter with a supermassive black hole. Theoretical pioneering work by \citet{Rees1990}  predicted 
   luminous flares of electromagnetic radiation from such events, lasting for weeks to months and then declining with a characteristic $t^{-5/3}$ law.
   A fraction of the disrupted star's matter is accreted onto the black hole (BH) while the rest is ejected. The accretion of the stellar material produces strong emission in different wavelength regimes. One of these is the X-ray regime which is a powerful tool for the detection of such events due to the huge peak brightness observed at these wavelengths. 
   TDEs were first detected in the X-ray band with the ROSAT mission \citep{KomossaBade1999, KomossaGreiner1999, Grupe1999, Greiner2000, Komossa2004} and are characterized by large peak luminosities up to $L_{\rm X}>10^{44}$ erg/s, super-soft X-ray spectra, high amplitudes of decline, and host galaxies which do not show AGN activity but are quiescent.  
   
   TDEs were then also identified at other wavebands including the radio, optical, UV and hard X-rays (review by \citealt{Komossa2015}). Only a small fraction of them were detected at radio frequencies and launches powerful jets \citep[e.g.,][]{Burrows2011, Bloom2011, Zauderer2011}. X-ray TDEs show a wide range of host galaxy and SMBH masses between $M_{\rm SMBH} \approx 10^{5-8}$ M$_{\odot}$ \citep{KomossaBade1999, Komossa2004, Maksym2013, Donato2014, Lin2017, Lin2021}, while optical and emission-line TDEs are preferentially detected in lower-mass hosts \citep{Komossa-et08, Wang2012, wevers17, Zhou2021}. 
   
   In the X-ray band, several new TDEs were identified with Chandra and XMM-Newton with peak luminosities up to a few times 10$^{44}$ erg/s and very soft X-ray spectra, located in quiescent host galaxies    \citep[e.g.,][]{Maksym2010, Esquej2008, rsaxton12, Maksym2013, Donato2014, Lin2015, Lin2017, Li2020, Lin2021}. Since TDEs are rare events, and their properties like light curve and spectral evolution can vary, identification of each single new event is of great interest. 

A detailed analysis of TDEs is important because it enables the investigation of matter under strong gravitational influence including precession effects in the Kerr metric \citep[e.g.,][]{stone12}. Furthermore, TDEs are a possible means to find intermediate mass BHs (e.g., \citealt{wevers17, Lin2017}), supermassive BH binaries \citep{Liu2014}, and recoiling BHs \citep{komossa08}. Additionally, one can draw conclusions about the spin of the black hole depending on the light curve of a TDE and the rate at which they occur \citep{kesden12}. Thus, their detailed study will greatly enhance our general understanding of BHs.

TDEs are best identified in {\em non-active} galaxies. Such galaxies lack the  high-ionization narrow emission lines which are very characteristic for AGN.  While AGN permanently harbor a (variable) accretion disk, in quiescent host galaxies such a disk is absent, and  luminous, giant-amplitude
X-ray flaring from the cores of quiescent galaxies can then be uniquely associated with TDEs \citep{rees88, KomossaBade1999}.

In this paper we present the identification and follow-up observations of a bright X-ray outburst.
In Section \ref{sec:analysis} we describe the analysis of its X-ray spatial, spectral and timing properties, as well as
optical follow-up spectroscopy and imaging. Finally, we discuss different outburst scenarios and conclude that the most likely mechanism to produce the event is a TDE (Sect. \ref{sec:discussion}).
When we report luminosities, these are based on a distance of 224 Mpc.

% Authors can give a citation as `\citealt{Michel+etal+1992}'.
% You may also use \cite, \citep and \citet for citation, and use Table~1
% or Figure~1 and so forth. Using \ref and \label for cross-references of
% Tables/Figures is a good way in adjusting/adding/removing text, tables or
% figures.

%\begin{figure} 
%   \centering
%   \includegraphics[width=12.0cm, angle=0]{sample2_fig1.eps}
%   \caption{ } 
%   \label{Fig1}
%   \end{figure}

%--------------------------------------------------------------------
\section{Observations}
\label{sec:observation}
The transient source RXJ133157.6--324319.7  presented in this work was detected in observations with the Roentgen Satellite (ROSAT, \citealt{truemper82}) of the galaxy cluster Abell 3560 performed in 1993 \citep{reiprich01}. For this observation, one of the Position Sensitive Proportional Counters (PSPC-B,  \citealt{briel86}) was used. A total of five observations over the course of 335 days covered this object. We determined the position of the outburst at RA: $13^\mathrm{h}31^\mathrm{m}57.66^\mathrm{s}$ and Dec: $-32^\circ43'19.7''$ by using the source detection implemented in the EXSAS software \citep{exsas}. 

\begin{figure}[ht]
   \centering
   \includegraphics{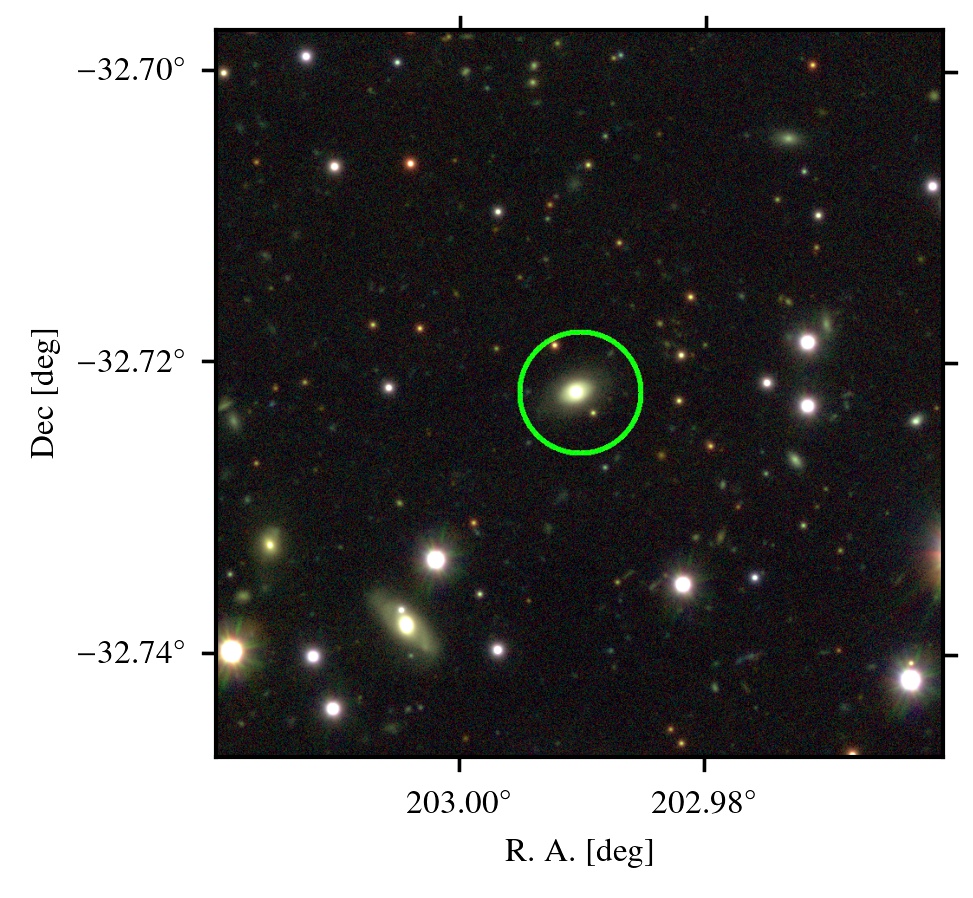}
   \caption{Optical image taken with OmegaCAM at the VST about 20 years after the X-ray outburst. The ROSAT PSPC error circle with a radius of $15''$  is overlaid.}
              \label{fig:opt_counter}%
    \end{figure}

By comparing the position with an optical image on the DSS and an image later taken with
the OmegaCAM 
at the VLT Survey Telescope \citep[VST;][]{arnaboldi98} in 2013, about 20 years after the outburst, one finds a galaxy to be the likely optical counterpart of the X-ray outburst. 
An image and spectrum of the galaxy was taken in 1999 (see Sect. 3.3). In  
Fig. \ref{fig:opt_counter} we overplot the X-ray error circle on the more recent galaxy image taken with OmegaCAM.

%-------------------------------------- Two column figure (place early!)

\section{Data Analysis}

\subsection{X-ray spectrum} 
\label{sect:X-ray_spectrum}

\begin{figure}[ht]
\hspace{-0.2cm}\includegraphics[width=0.7\textwidth,viewport=78 50 570 405, clip, angle=270 ]{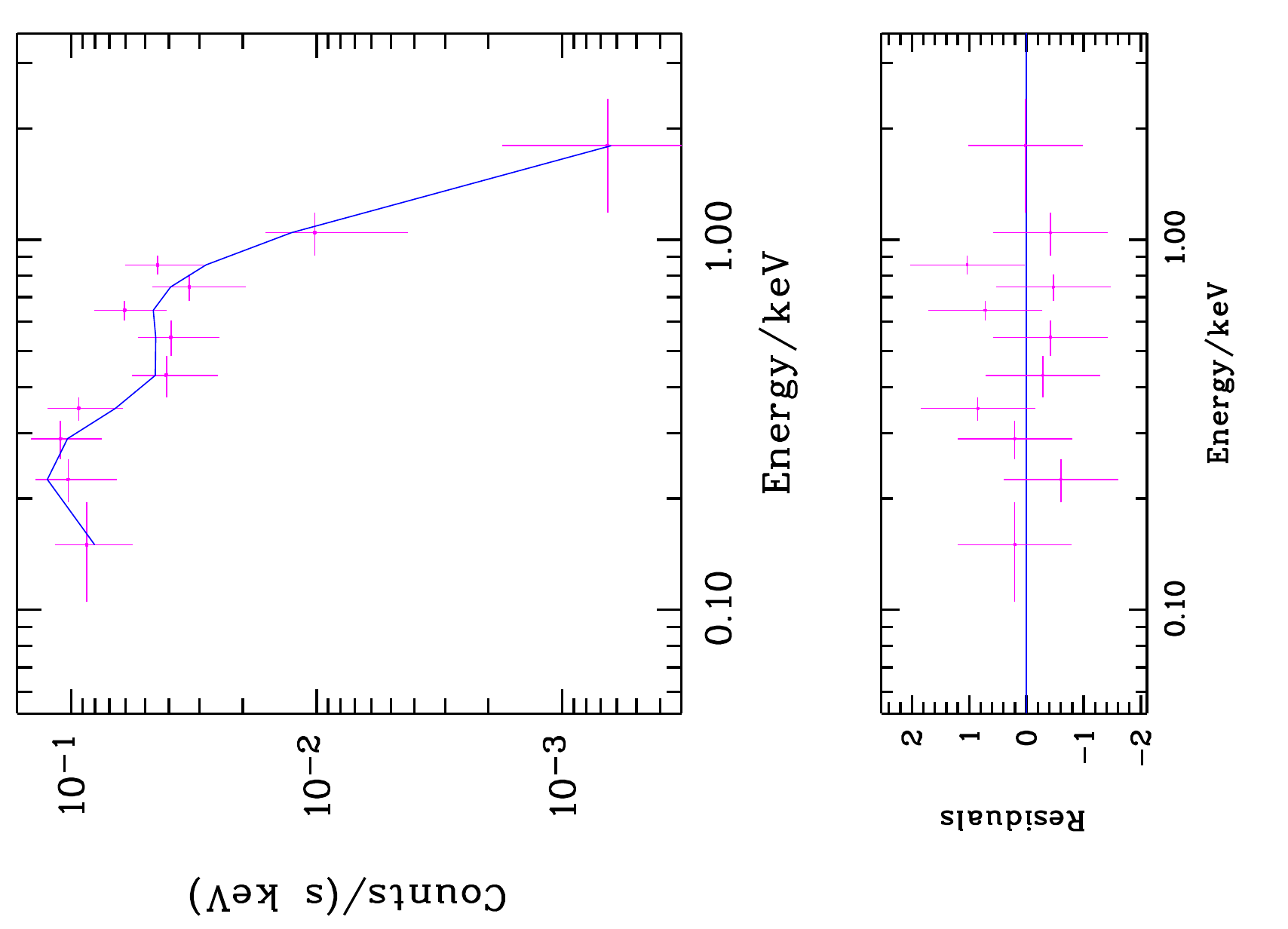}
\caption{Spectrum of the source in part 1 of observation 800284p  in the energy range $0.1\too 2.4\,\mathrm{keV}$, fit with a blackbody model. This is the individual observation with the largest number of counts, though not the peak of the light curve (see Tab. \ref{tab:data}).
}
\label{fig:spectrum}
\end{figure}

All ROSAT X-ray analysis was done with the EXSAS software \citep{exsas}.
{A spectrum of the source was extracted for part 1 of observation 800284p, taken on 25-01-1993 (Fig. \ref{fig:spectrum}) which has the largest number of counts.
After vignetting correction, we fit a simple power law model to the background-subtracted spectrum. However,  this  results in an unusually steep slope with photon index 
$\Gamma_{\rm x} = -5.87\pm{1.80}$
and is not discussed further.
Next, a black body model was fit to the spectrum, as successfully applied to all previous soft ROSAT TDE spectra \citep[e.g.,][]{KomossaBade1999}. 
Such a model fits the spectrum well. We find $N_H = (0.29\pm0.29) \times 10^{21}$ consistent with Galactic absorption ($0.38 \times 10^{21}$, \citealt{2016A&A...594A.116H}), 
$kT$ = 105 $\pm$ 31 eV and $\chi^2_{\rm red}$=1.1. 
The inferred}
unabsorbed X-ray luminosity for this time interval is  $L_X(0.1-2.4\,\mathrm{keV}) = 6.05 \times 10^{42}\,\mathrm{erg\,s}^{-1}$. 

\begin{figure}
\centering
\includegraphics[width=0.8\textwidth,clip,]{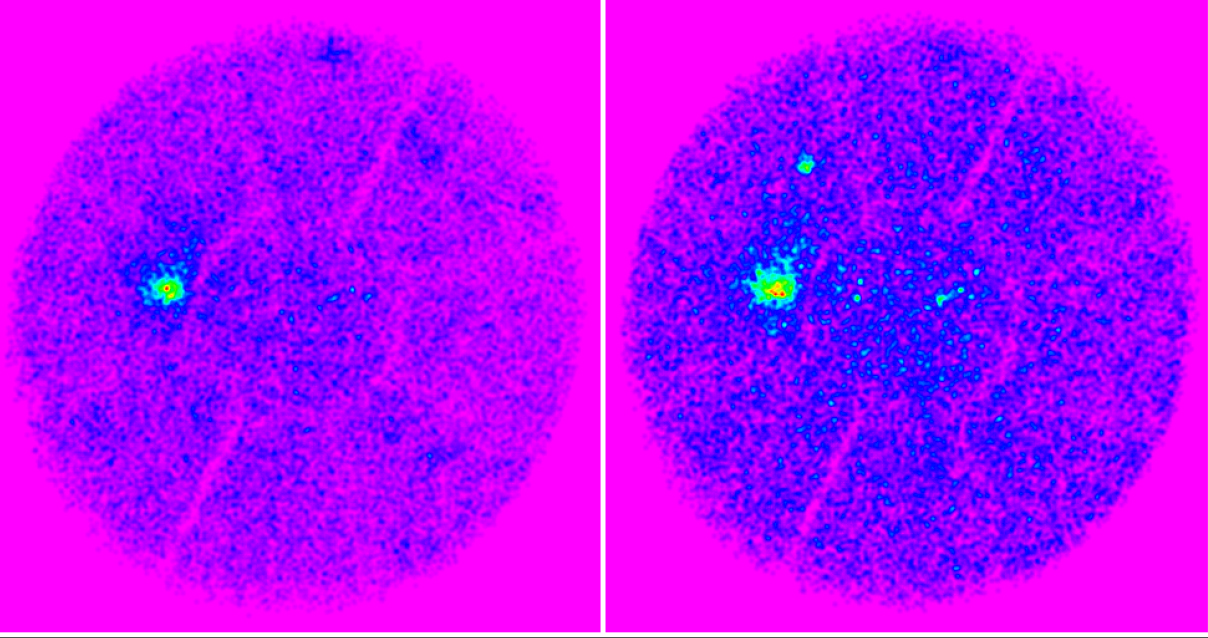}
\caption{X-ray photon image of the ROSAT PSPC field of view of Abell 3560 without (left, observation 800284p-1) and with (right, observation 800284p) the X-ray transient source RXJ133157.6–324319.7. The PSPC field of view has 2 degrees diameter.}
\label{fig:X-ima}
\end{figure}

An X-ray image of the field of view including the cluster Abell 3560 with and without the TDE is displayed in Fig. \ref{fig:X-ima}. 

\label{sec:analysis}
\subsection{Light curve}

{Based on further PSPC observations with fewer photons, but still enough to determine countrates or upper limits,
(Obs.~IDs 800381p, 800381p-1, 800284p, 800284p-1, and 800381p-2), the event light curve was constructed.  }

\begin{figure*}[ht]
\includegraphics[width=0.85\textwidth]{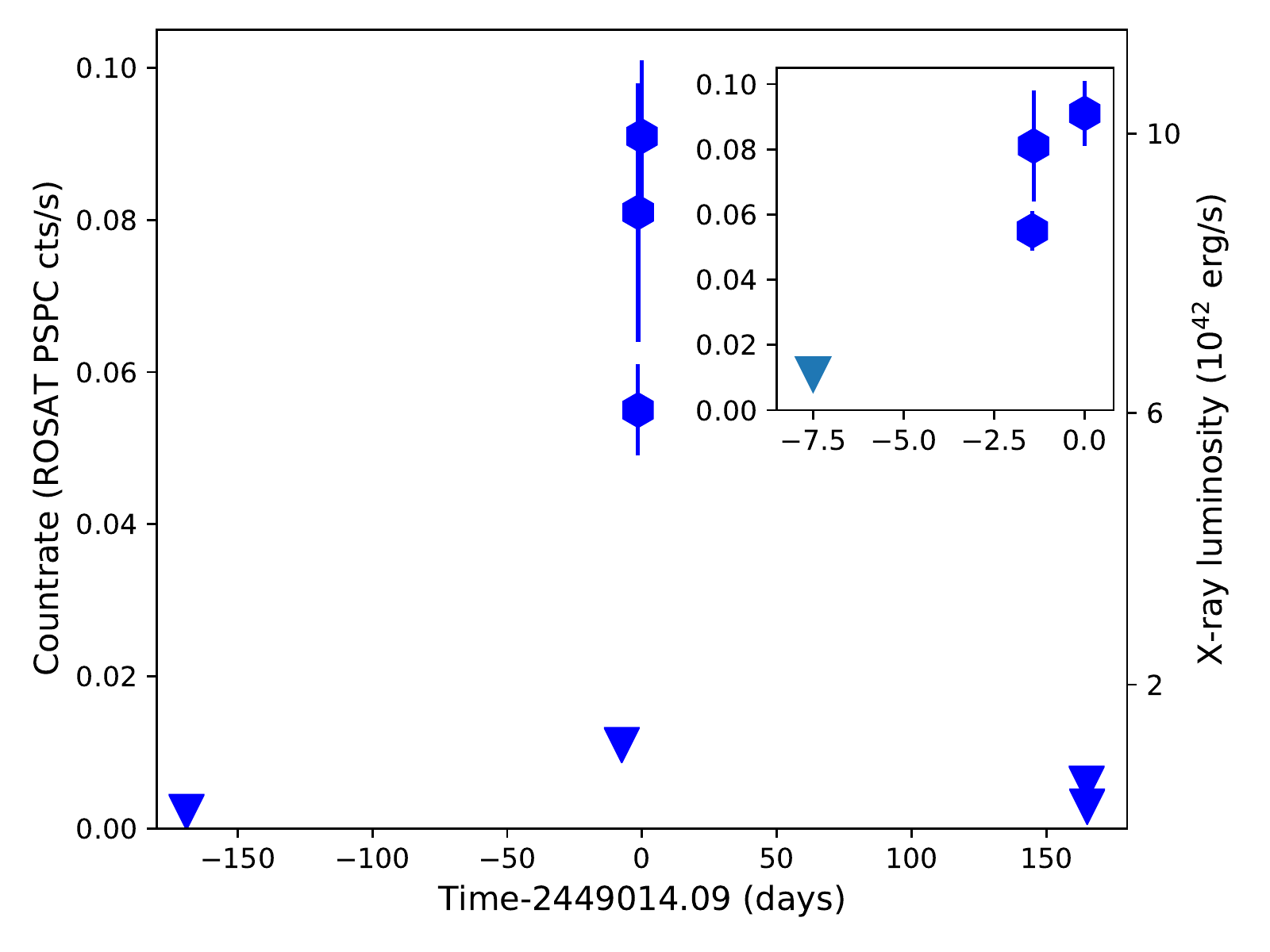}
\caption{Light curve of RXJ133157.6--324319.7 in the (0.1--2.4) keV band. The zero time corresponds to the time of maximum brightness (2449014.09 JD). The left y-axis labelling is for the observed rate, while the right one is based on the spectral fit of the first part of observation 800284p and under the assumption that the spectrum did not change over time. The symbol size is wider than the corresponding time bins.
\label{fig:luminosity}}
\end{figure*}

Count rates in the energy band (0.1--2.4) keV are measured in an area with a radius of $165''$ around the source and corrected for dead time, exposure time, and vignetting effects. A background subtraction using an annulus in the radial range  $300''$--$450''$ with no sources is carried out. The exact choice of background location does not change the result significantly. The results are listed in Tab.~\ref{tab:data}.
In Fig. \ref{fig:luminosity} the light curve is shown.
The luminosities all assume the same spectrum (Section~\ref{sect:X-ray_spectrum}) and were derived using the redshift $z=0.051$ we measured form the optical spectrum (Sect. 3.3), and using a distance of 224 Mpc. 

Comparing the  countrate at maximum, 0.091 cts/s,  to the upper limit 170 days before, $2.2\times 10^{-3}$ cts/s, and 165 days after, $2.9\times 10^{-3}$ cts/s, an increase and decrease by factors of $>$41 and $>$31, respectively, are implied, each with a $\sim$ $10\%$ 
uncertainty. 
Potentially even more interesting is the fast rise within 8 days of a factor of $>$ 8, based on the lower limit 0.011 cts/s.

\begin{table*}[!ht]
\caption{Data extracted from five ROSAT PSPC observations between August 1992 and July 1993. The "p-1" after an Obs-ID indicates splitting due to the ROSAT scheduling. The data of 800381p-1 and 800284 were split into two time slices each (called pt1 and pt2) according to data gaps.  The on-axis exposure (column 3) cannot be combined with the observed  number of counts (5th col) to compute the count rate (4th col, corrected to reflect an on-axis count rate) due to the vignetting at the off-axis position of the source. All numbers are for the (0.1--2.4) keV band.}
\centering
\begin{tabular}{lccccc}
\hline
Observation ID & Mid-Time of Obs. & Exposure & Source Count Rate  & Source Counts & $L_{\rm X}$\\
        & (JD) & (s) & (PSPC cts/s) & (cts) & $10^{43}$ erg/s\\
\hline
800381p    & 2448845.0983 & 3612 & $<$2.2E-3 & -- & $<$0.024\\
800381p-1 pt1& 2449006.5937 & 1932 & $<$0.011 & -- & $<$0.121\\
800284p pt1 &  2449012.6458 & 3559 & 0.055$\pm$0.006 &  131$\pm$15 & 0.60$\pm$0.07\\
800284p pt2 &  2449012.6805 &  ~~541 & 0.081$\pm$0.017 & 30$\pm$6 & 0.89$\pm$0.19\\
800381p-1 pt2 & 2449014.0906  & 1076 & 0.091$\pm$0.010 & 80$\pm$9 & 1.00$\pm$0.11\\
800284p-1  & 2449179.0550 & 3060 & $<$5.9E-3 & -- & $<$0.065\\
800381p-2  &2449179.2655 &5796 &    $<$2.9E-3 & -- & $<$0.032\\
\hline
\label{tab:data}
\end{tabular}
\end{table*}

Furthermore, we used the Neil Gehrels Swift Observatory \citep[Swift hereafter;][]{Gehrels2004} to search for late-time X-ray emission from RXJ133157.6--324319.7. We determined upper limits (2$\sigma$) from these three Swift XRT observations performed in April 2018 (Obs-IDs: 00010661001, 00010661002, and 00010661003). This resulted in  limits consistent with those from the ROSAT PSPC pointed observations taken about 25 years earlier, again assuming the same spectrum (Tab.~\ref{tab:upplim}). 
Since the three upper limits determined with Swift are consistent with each other, we have also merged the three single observations to determine one deeper upper limit of $0.001$ cts/s.
Limits from the ROSAT
All-Sky Survey and XMM-Newton Slew Survey are also shown in the table.

\begin{table}[ht]
    \centering
    \caption{Upper limits at 2$\sigma$ confidence from other X-ray observations.}
    \setlength{\tabcolsep}{0.3em}
    \begin{tabular}{lcccc}
    \hline
    \noalign{\smallskip}
    Mission & Date & UL     & exposure & $L_{\rm X}$ \\
            &    & (cts/s)& (s)    & (10$^{43}$ erg/s) \\
    \noalign{\smallskip}
    \hline
    \noalign{\smallskip}
    ROSAT Survey & 1991/01/04 & $<$0.027 & 289 &  $<$0.297 \\
    XMM slew & 2006/01/10 & $<$0.36 & 10 &   $<$0.412  \\
    XMM slew & 2015/02/03 & $<$0.50 & 7 &  $<$0.572  \\
    XMM slew & 2015/08/07 & $<$0.75 & 5 & $<$0.858\\
    Swift & 2018/04/18 & $<$0.0033 & 1123 & $<$0.101 \\
    Swift & 2018/04/22 & $<$0.0020 & 1971 & $<$0.061\\  
    Swift & 2018/04/26 & $<$0.0029 & 1998 & $<$0.088\\
    \noalign{\smallskip}
    \hline
    \end{tabular}
    \label{tab:upplim}
\end{table}

\subsection{Optical observations}
\label{optical_observation}
A 300 s I band image, taken on January 26, 1999 with DFOSC
at the 1.5m Danish telescope at ESO/La Silla using a 2052x2052 backside
illuminated LORAL/LESSER chip shows an extended galaxy within the X-ray error circle, as well as
a few further, much fainter star-like objects.

\begin{figure*}
   \centering
\includegraphics[width=0.95\textwidth]{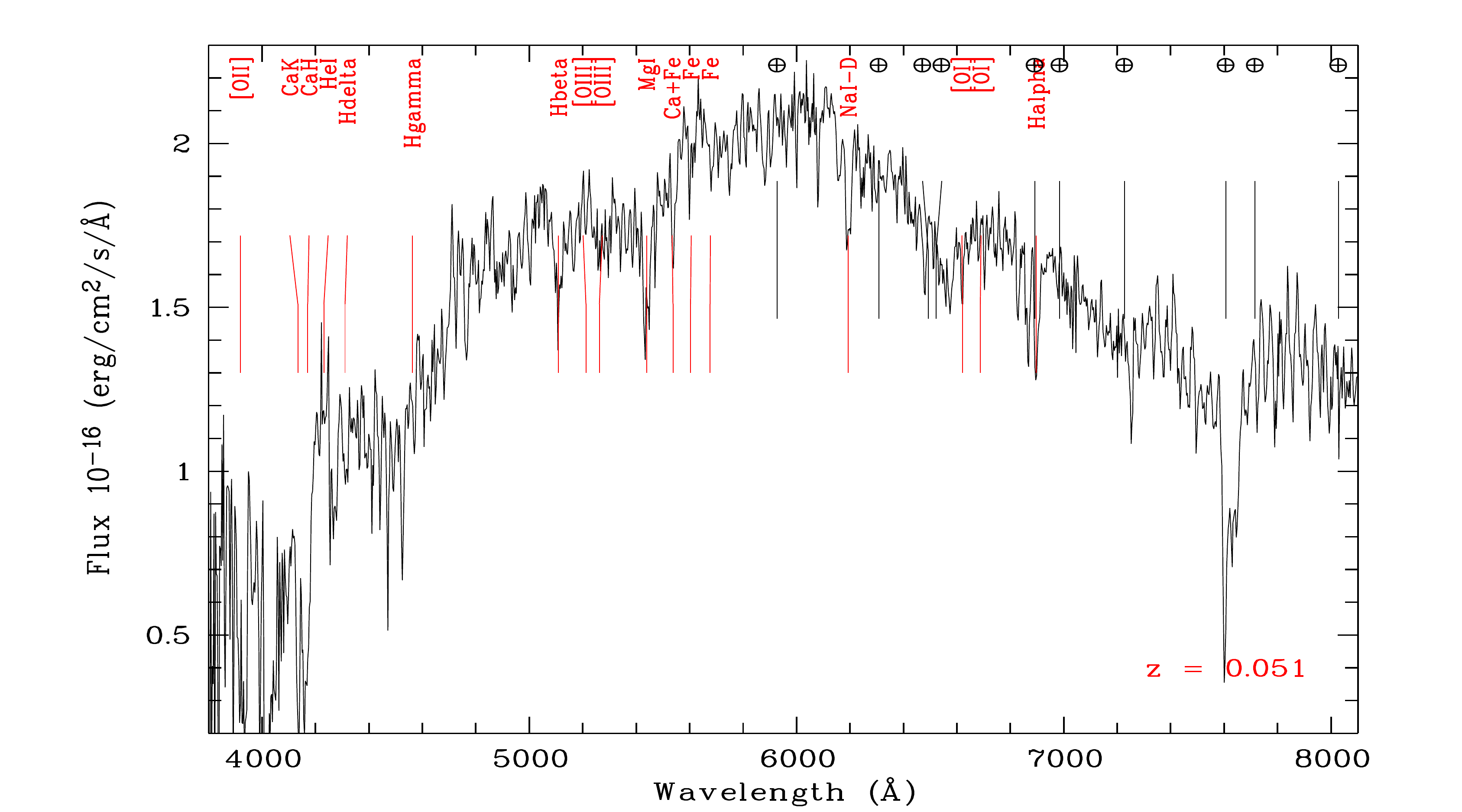}
   \caption{Optical spectrum of the galaxy at the position of the X-ray transient RXJ133157.6--324319.7, 
taken with the Danish 1.5m telescope on La Silla about six years after the outburst.  
Prominent absorption lines (detected) and emission lines (undetected) are labelled in red. 
Crosses mark the location of atmospheric absorption.   
 }  
 \label{fig:opt_spektrum}%
    \end{figure*}

The bright central galaxy, with coordinates RA (2000) = $13^\mathrm{h}31^\mathrm{m}58.2^\mathrm{s}$ and 
Dec (2000) = $-32^\circ,43'20''$,  
was observed on January 26, 1999 with DFOSC using grism 4 with 300 grooves
per mm, covering the 3000--9000 \AA\ range at a dispersion of
220 \AA/mm or 3.0 \AA/pixel. The seeing was $1.4''$, giving a FWHM
resolution of 11 \AA. Two exposures of 1500 s and 1800 s were taken, 
respectively. Standard processing and optimal extraction were done using
canonical MIDAS routines. The star GD 108 has been used for flux calibration.
The averaged spectrum is shown in Fig.~\ref{fig:opt_spektrum}, together with some major absorption 
lines. We determine R $\approx$ 17.7 mag from the spectrum.

The optical spectrum is characterized by strong absorption
lines of Na\,I 5175\AA, Mg\,I 5890\AA~and H$\beta$ (note that
H$\alpha$ overlaps with the atmospheric B band), typical of 
an elliptical or early spiral type. Using the Mg\,I, Na\,I and H$\beta$ 
line we derive a redshift of $z=0.051\pm0.001$, similar to the redshift
of Abell 3560 ($z=0.0495$, \citealt{2002ApJ...567..716R}). The deep Balmer lines
and the strong drop of the flux beyond the Ca\,II H/K break argue 
against a classification as BL Lac object. No AGN-like forbidden 
emission lines, like [OIII]$\lambda$5007, are detected. These observations establish a quiescent, non-active host galaxy.

We also retrieved public OmegaCAM@VST
data around the cluster Abell 3560 in the filters $i_{\rm Sloan}$,
$r_{\rm Sloan}$, and $g_{\rm Sloan}$. The data were obtained under
programs 091.A-0050(F) for the $r$-band, 094.A-0050(A), 092.A-0057(D)
and 092.B-0623(D) for the $g$-band and 089.A-0095(H) for the $i$-band.
All images were obtained under very good, sub-arcsecond and
photometric observing conditions. The total exposure times are 3360
s ($g$-band), 2368 s ($r$-band) and 1000 s
($i$-band). The data were processed with the THELI-pipeline \citep[see][]{Erben05} and the processing methods are described in detail in
\citet{kuijken15}. Fig.~\ref{fig:opt_counter} shows a 6 arcminute by 6 arcminute cutout
around the outburst position of RA: $13^\mathrm{h}31^\mathrm{m}57.66^\mathrm{s}$ and Dec: $-32^\circ43'19.7''$. The ROSAT error circle is displayed and corresponds to $15''$. At the position we
identify a bright object which we visually identify as a regular
elliptical galaxy, in agreement with the spectroscopic analysis, presumably a member of the Abell 3560 cluster given the coincident spectroscopic redshift determined above.
The magnitudes of the galaxy are $i_{\rm Sloan}\approx 16.6$, $r_{\rm
Sloan}\approx 16.9$ and $g_{\rm Sloan}\approx 17.7$.

\section{Discussion}
\label{sec:discussion}
\subsection{Likelihood of TDE}

\begin{sloppypar}The X-ray observations of a transient source fit the criteria of a tidal disruption event, as first predicted by theory and then observed in the X-ray regime, especially with ROSAT in the same energy band as discussed here 
\citep{rees88, KomossaBade1999}
A large change in brightness by at least a factor of 40 leading to a  luminosity of at least \hbox{$(1.0 \pm 0.1)\times 10^{43}\,\mathrm{erg/s}$} in the 0.1--2.4 keV energy band is detected. The position of the outburst is coincident within the errors with the nucleus of a quiescent galaxy in the galaxy cluster Abell 3560. An expected soft X-ray spectrum can be confirmed with a
black body temperature of $kT$ = 105 $\pm$ 31 eV.
In Section \ref{sec:alternativeOS} alternative outburst scenarios will be discussed and are found to be unlikely or ruled out. 
\end{sloppypar}
This suggests the X-ray emission to originate from a TDE. The light curve of the event is not covered well. 
However, the event fulfills the expectations for an X-ray TDE because it shows a quiescent, non-active host galaxy which rules out AGN activity. Further, it shows no evidence of large scale jets, no reoccurring X-ray emission, but a rapid increase and then a large decrease.
Also 25 years after the outburst we could only determine an X-ray upper limit with Swift consistent with the ROSAT upper limit half a year after the outburst.

\subsection{Alternative Outburst Scenarios} 
\label{sec:alternativeOS}
\subsubsection{AGN}
A source that could potentially be mistaken for a TDE is an AGN. However, our optical spectrum clearly establishes a non-active, quiescent host galaxy, since the characteristic narrow emission-lines of AGN are undetected, and since the source is not a blazar either.
Furthermore, as already mentioned above, all properties of the event show the characteristics of a TDE, consistent with the systems previously observed.  

\subsubsection{Supernova and X-ray Binary in the Quiescent Galaxy}
A high fluctuation in luminosity like the one observed could potentially be explained by supernova explosions. However, the peak brightness is unusually high for a supernova. Their typical luminosities range from $10^{37}$ to $10^{40}\,\mathrm{erg\,s}^{-1}$ and they very rarely reach up to $10^{42}\,\mathrm{erg\,s}^{-1}$ \citep{dwarkadas12} which is still an order of magnitude lower than detected in this observation. A similar argument can be made to exclude X-ray binaries in the quiescent galaxy as counterpart of the X-ray outburst. While large fluctuations in X-ray emission on this time scale could arise from X-ray binaries, their X-ray luminosity is expected to be much lower than the highest luminosity observed here. A recent study of 110 X-ray outbursts in 36 low mass X-ray binaries performed by \citet{yan15} show that the typical peak luminosities only range from $10^{37}\too 10^{38}\,\mathrm{erg\,s}^{-1}$. Additionally, black hole X-ray binaries are not expected to show such soft spectra. In the soft state power law indices from
%\hbox{$-2\too -3$}
$-2$ to $-3$
are expected \citep{gilfanov10}. \citet{sazonov17} find similar results for high mass X-ray binaries.

\subsubsection{X-ray Binary in the Milky Way}
The derived luminosity depends on the distance of the source and was determined under the assumption, that the source is located in the member of the galaxy cluster Abell 3560. However, if the X-ray outburst actually occurred much closer to us and it only is due to a coincidental projection effect that its position overlaps with the optical location of the galaxy in Abell 3560, the peak luminosity has been overestimated. Instead, if one assumes that the X-ray source originates from within the Milky Way (Galactic coordinates: $l=312.7^\circ$ and $b=29.4^\circ$), the luminosity is estimated by $L=4\pi SD^2$ where $S$ is the derived flux and $D$ the distance to the source. Results by \citet{Coleiro13} suggest that only few detected X-ray binaries exceed a distance of $10\,\mathrm{kpc}$ while many are found to be at a distance of about $2$--$3\,\mathrm{kpc}$. Then, the resulting (peak) luminosity is about ten orders of magnitude lower, $\sim$$1\times 10^{33}\,\mathrm{erg\,s}^{-1}$.

This is quite low for a typical observed Milky Way X-ray binary but does not rule out this origin; also, the column density might be lower than what we assumed here, especially for a smaller distance. However, the very soft spectrum ($kT=0.1$ keV)
makes an X-ray binary origin in our Galaxy unlikely unless their spectra soften significantly at these not very well studied, very low luminosities.

\subsubsection{Gamma-ray Burst}

Another alternative scenario to explain an X-ray outburst is a gamma-ray burst (GRB) afterglow. Comparing the time and position of the event with detected gamma-ray bursts listed in the Gamma-Ray Bursts Catalogue (GRBCAT) results in only one match in the vicinity of the X-ray detection at the time of the outburst: GRB 930118 \citep{grb}. It is the only event that occurred during a time interval from $2448997.5$ to $2449018.5\,\mathrm{JD}$
%Jan 10-31
and within a maximum distance of $30^\circ$ relative to the position of the transient. While the detection overlaps with the beginning of the rise of the X-ray detection, GRB 930118 occurs seven days prior the detected peak luminosity. 
It was detected by the Burst And Transient Source Experiment (BATSE) and its position was refined with the Imaging Compton Telescope (COMPTEL) to be at RA: $14^\mathrm{h}47^\mathrm{m}12^\mathrm{s}$ and Dec: $-34^\circ48'00''$ with an uncertainty of $\sigma=1.5^\circ$ \citep{grb}. The two events are, therefore, separated by 15.7 degrees, 
making it impossible that the same event was the origin of both detections. Moreover, the soft spectrum is atypical for GRB afterglows,
arguing against an afterglow interpretation from a burst not detected by any active GRB satellite mission.

\subsection{Rise Time}

A rise in brightness within 8 days was observed. While
the majority of X-ray TDEs were only observed after their peak (or months before their peak), the case of NGC 5905 detected with ROSAT showed a rise by a factor of $\sim3$ during a similar one week time interval \citep{Bade1996, KomossaBade1999}. We discuss several scenarios for the fast variability. 

\citet{guillochon13} show in simulations that rise times are expected to range from 24 to 32 days, depending on the type of star. \citet{lodato09} show numerically that slightly faster rise times can be achieved, depending on the distribution of the matter in the star. The more homogeneously distributed the matter is, the faster the rise time. 
It is well possible that the rise we observed within those 8 days is not the total time of the rise, as other systems showed higher peak luminosities \citep{Komossa2015}. 
 However,  it is also well possible, that we did not observe the actual rise itself, but rather saw fluctuation in X-ray luminosity during rise or decline. 
 A few TDEs with large short-timescale fluctuations in luminosity have been observed, and the underlying mechanism may also be operating in the event presented here, as discussed below.

One possible explanation is high-amplitude fluctuations in the light curve (so we do not observe the actual rise time). Strong continuous fluctuations on very short time scales have previously been observed in jetted TDEs \citep[e.g.,][]{Zauderer2011, Burrows2011, csaxton12}
where for instance beaming could drive the observed rapid X-ray emission
(see also \citep{Wong2007}).
Some of these fluctuations happen on time scales as short as $100\,\mathrm{s}$. However, for the presented TDE, no matching radio emission has been found when checking the NASA/IPAC extragalactic database (NED) for the corresponding coordinates; although, there may not have been radio observations in the relevant time range. 

A second possibility is that the TDE originated in a binary SMBH system. Under such circumstances, the second SMBH temporarily interrupts the accretion stream on the primary, leading to characteristic fast dips in TDE light curves, and then excess emission at the times the primary starts accreting again. This process has been identified in the light curve of SDSSJ1201+30 \citep{Liu2014} and a similar mechanism could be at work in the system presented here.  A better light curve coverage would have been needed to constrain this scenario further. 

Another possibility is that the mass of the black hole is actually smaller than the typical mass of SMBHs in the center of galaxies. If instead of a mass of $10^6\msun$ a black hole mass of \hbox{$6\times 10^4 \too 10^5\msun$} was assumed, the X-ray outburst could be explained by the disruption of a main sequence star \citep[based on][]{guillochon13}. These kind of black holes could potentially be found in dwarf galaxies surrounding the detected galaxy \citep[e.g.,][]{reines13}. However, if the TDE originated in a dwarf galaxy with a SMBH mass as low as \hbox{$6\times 10^4\msun$}, the high observed peak luminosity would be unexplained, especially if we did not catch the event right at peak.

\section{Conclusions}

We present a new X-ray selected TDE candidate, RXJ133157.6--324319.7, exhibiting a fast
variability
time and high peak luminosity. The data show properties which are expected for a typical tidal disruption event, observed in the majority of the previously identified X-ray events:
\begin{enumerate}
\item The occurrence of one outburst in the X-ray regime at the location of an optically quiet galaxy ($z=0.051$). Additionally, no signs of reoccurring X-ray emission were found. 

\item The spectrum is very soft and is well described by a black body model of $kT=0.1$ keV; a value very similar to other ROSAT soft X-ray TDEs. No excess absorption is required.

\item The highest observed luminosity is $1\times 10^{43}$ erg/s.

\item An increase in luminosity by a factor 8
is observed within 8 days.  
No X-ray emission is detected 165 d after the peak
(implying a factor $>$40 decline), and none is detected with Swift in 2018. 
% See above.
\end{enumerate} 
These observed quantities fit the expectations for a TDE very well.
Other outburst scenarios can be ruled out or seem to be unlikely:
\begin{enumerate}
\item An AGN can be ruled out 
because of the {\em quiescent} host galaxy with no characteristic optical narrow emission lines detected at all. 
\item A supernova or X-ray binary in the quiescent galaxy as a source seems to be highly unlikely because of the high luminosity. The X-ray luminosities of these events are smaller than the one of the detected source by one to several orders of magnitude. 
\item It does not appear to be the afterglow of a gamma-ray burst, either. No burst has been detected nearby the determined position of the outburst in the relevant time interval. A neutron star-neutron star or neutron star-black hole merger with very weak gamma-ray emission but very strong X-ray emission also seems unlikely.
\item We cannot entirely rule out an  optically faint X-ray binary in our galaxy, projected by chance right onto the galaxy at $z=0.051$. However, the soft spectrum, and missing recurrence of X-ray emission speak against this scenario. 
\end{enumerate}

A possible explanation for the factor of 8 rise within 8 days is a large fluctuation in brightness as has been found in a small number of other TDEs  (so we do not see the actual rise time) as it could be produced, for instance, in a binary SMBH system.
Alternatively, the mass of the accreting black hole could be significantly lower than for typical supermassive black holes. 

Further observations with current X-ray telescopes would enable more accurate upper limits on, or a detection of, the baseline emission and, therefore, a better estimation of the total amplitude of variability and the spectrum and nature of the low-state emission, if any. 

\normalem
\begin{acknowledgements}
We would like to thank Julia Hampel  for carrying out her master thesis in Bonn, in collaboration with the co-authors.
This work is based on an initial paper draft of hers which she could not quite complete due to her sudden, unexpected,
and very saddening death. We would also like to thank Melanie Hampel for help with locating relevant
files. 
Furthermore, we would like to acknowledge our referee for constructive comments.
This work is based on data taken with the X-ray satellites ROSAT, XMM-Newton and Swift, and on data taken with the optical 1.5m Danish telescope and OmegaCAM. 
The ROSAT project  was supported by the German
Bundesministerium fur Bildung, Wissenschaft, Forschung und Technologie (BMBF/DARA) and the Max-Planck-Society.
XMM-Newton is an ESA science mission with instruments and contributions directly funded by ESA Member States and NASA.
This research has made use of the NASA/IPAC Extragalactic Database (NED) which is operated by the Jet Propulsion Laboratory, California Institute of Technology, under contract with the National Aeronautics and Space Administration.
\end{acknowledgements}
  
\bibliographystyle{raa}
\bibliography{bibtex}

\end{document}